\documentclass[sigconf,x11names]{acmart}

\AtBeginDocument{%
  }

\copyrightyear{2025}
\acmYear{2025}
\setcopyright{rightsretained}
\acmConference[ASPDAC '25]{30th Asia and South Pacific Design Automation Conference}{January 20--23, 2025}{Tokyo, Japan}
\acmBooktitle{30th Asia and South Pacific Design Automation Conference (ASPDAC '25), January 20--23, 2025, Tokyo, Japan}
\acmDOI{10.1145/3658617.3697687}
\acmISBN{979-8-4007-0635-6/25/01}

\usepackage{listings}
\usepackage{graphicx}
\usepackage{color}
\usepackage{xcolor}
\usepackage{url}
\usepackage{hyperref}
\usepackage{multirow} 
\usepackage{booktabs} 
\usepackage{subcaption}
\usepackage{caption}
\usepackage{colortbl}
\usepackage{algorithmic}
\usepackage{algorithm}

\definecolor{mauve}{rgb}{0.58,0,0.82}
\definecolor{dkgreen}{rgb}{0,0.6,0}
\definecolor{gray}{rgb}{0.7,0.7,0.7} 
\newcommand{\grayline}{\arrayrulecolor{gray}\hline\arrayrulecolor{black}}

\newcommand{\N}{LUTMUL}
\newcommand{\eg}{\textit{e}.\textit{g}.}

\begin{document}

\title{LUTMUL: Exceed Conventional FPGA Roofline Limit by \underline{LUT}-based Efficient \underline{MUL}tiplication for Neural Network Inference}


\author{Yanyue Xie}
\affiliation{%
  \institution{Northeastern University}
  \country{}}
\email{xie.yany@northeastern.edu}

\author{Zhengang Li}
\affiliation{%
  \institution{Adobe}
  \country{}}
\email{li.zhen@northeastern.edu}

\author{Dana Diaconu}
\affiliation{%
  \institution{Northeastern University}
  \country{}}
\email{diaconu.d@northeastern.edu}

\author{Suranga Handagala}
\affiliation{%
  \institution{Northeastern University}
  \country{}}
\email{s.handagala@northeastern.edu}

\author{Miriam Leeser}
\affiliation{%
  \institution{Northeastern University}
  \country{}}
\email{mel@coe.neu.edu}

\author{Xue Lin}
\affiliation{%
  \institution{Northeastern University}
  \country{}}
\email{xue.lin@northeastern.edu}


\renewcommand{\shortauthors}{Yanyue Xie, Zhengang Li, Dana Diaconu, Suranga Handagala, Miriam Leeser, and Xue Lin}

\begin{abstract}
For FPGA-based neural network accelerators, digital signal processing (DSP) blocks have traditionally been the cornerstone for handling multiplications. This paper introduces \texttt{\N{}}, which harnesses the potential of look-up tables (\underline{LUT}s) for performing \underline{mul}tiplications. The availability of LUTs typically outnumbers that of DSPs by a factor of 100, offering a significant computational advantage. By exploiting this advantage of LUTs, our method demonstrates a potential boost in the performance of FPGA-based neural network accelerators with a reconfigurable dataflow architecture. Our approach challenges the conventional peak performance on DSP-based accelerators and sets a new benchmark for efficient neural network inference on FPGAs. Experimental results demonstrate that our design achieves the best inference speed among all FPGA-based accelerators, achieving a throughput of 1627 images per second and maintaining a top-1 accuracy of 70.95\% on the ImageNet dataset. 
\end{abstract}



\maketitle

\section{Introduction}
\label{sec:introduction}

Field-Programmable Gate Arrays (FPGAs) have been widely used as deep learning accelerators, facilitating advancements in computer vision~\cite{zhang2015optimizing, dong2023heatvit, li2024quasar, yang2024sda} and natural language processing~\cite{li2020ftrans, hong2022dfx, zeng2024flightllm, chen2024understanding} tasks. However, FPGAs lag behind Graphics Processing Units (GPUs) in terms of performance and ease of programming. FPGA reconfigurable logic mainly consists of look-up tables (LUT), block RAMs (BRAMs), and digital signal processing (DSP) blocks. Together with routing resources, FPGA can be reconfigured for customized designs. Despite the flexibility, FPGAs face constraints in clock frequency, floating-point performance, and memory bandwidth. This performance gap between FPGAs and GPUs is becoming even larger when considering the tensor core performance of GPUs. To address this, we need an algorithm-hardware co-design method to boost FPGAs with greater inference capability.

FPGA accelerators can follow GPU-like~\cite{wu2019high,yan2021fpga} architecture, which maps computation to compute cores with repetitive use. While beneficial, this approach encounters memory bandwidth issues similar to GPUs. Compared with GPUs, FPGAs usually have lower memory bandwidth, and the lower clock frequency of FPGAs means a lower upper bound of performance. While FPGA-based accelerators with specific instruction set architectures~\cite{yu2020light} offer flexibility across different models, they often compromise on performance due to non-optimized compute kernels for specific neural network layers.

To bridge the performance gap between FPGAs and GPUs, particularly in deep learning applications, we introduce \texttt{\N{}}, which leverages the look-up tables on FPGAs for deep learning tasks, focusing on accelerating convolutional neural networks (CNNs). We recognize that the traditional FPGA designs, heavily dependent on DSP blocks, may not fully exploit the parallelism and flexibility that LUTs offer, as the availability of LUTs typically outnumbers DSPs by a factor of 100. Our method emphasizes a novel utilization of LUTs to enhance computational efficiency and throughput in deep learning applications. Specifically, we embed the convolutional neural network weights into LUTs, where the LUT input is the activations and the LUT output is the multiplication result. Different from LUT-based general multipliers, our method is efficient in resources (requiring just 2 LUTs for a single 4-bit multiplication) and helps fully exploit the parallelism.

We propose a reconfigurable dataflow architecture for our LUT-based efficient multiplication kernel. Our dataflow architecture minimizes the memory access time by processing the data on-chip through each layer without external memory. The reconfigurability of the FPGA allows us to tailor the architecture specifically for each distinct layer of CNNs. With LUT resources, the generated accelerator can potentially exceed the peak performance of conventional DSP-based FPGA accelerators. Our dataflow architecture aims to enhance the overall efficiency of our deep learning accelerators, optimizing FPGAs for deep learning tasks.

Our contributions can be summarized as follows:
\begin{itemize}
    \item We present \texttt{\N{}}, an algorithm-hardware co-design method that embeds quantized neural network weights into look-up tables for efficient multiplications and uses dedicated look-up tables for full parallelism.
    \item We design a reconfigurable dataflow architecture that exploits scalability and LUT potential to save computational resources.
    \item Using \texttt{\N{}}, FPGA designs can potentially exceed the peak performance of conventional DSP-based FPGA accelerators when using the same amount of resources.
\end{itemize}
\section{Background}
\label{sec:background}

\subsection{Roofline Model Analysis}

GPUs leverage Single Instruction Multiple Data (SIMD) architecture, allowing them to simultaneously perform the same operation across multiple data points. This parallelism makes GPUs exceptionally efficient for tasks that can be divided into smaller, similar operations, such as matrix multiplication in deep learning.

FPGAs, by contrast, achieve parallel processing through their reconfigurability, allowing hardware to be tailored to specific computational tasks. This flexibility allows FPGAs to efficiently handle complex and diverse data processing tasks, offering advantages over the fixed architecture of GPUs. While FPGAs lack the raw SIMD power of GPUs for certain applications, they excel in scenarios requiring custom hardware configurations or low-latency, such as specific signal processing tasks or custom machine learning algorithms. However, this adaptability often comes with a trade-off in processing speed and ease of programming, with FPGAs typically lagging behind the computational throughput of GPUs.

The roofline model~\cite{williams2009roofline} is a useful tool for analyzing the performance of both GPUs and FPGAs. An algorithm running on GPUs or FPGAs can be either compute bound or memory bound. According to the roofline model~\cite{zhang2015optimizing}, the peak performance of FPGAs is:
\begin{equation}
    Peak\ performance = p \times PEs \times 2 \times f,
\end{equation}
where $PEs$ is the number of processing elements (PEs) used in the accelerator, such as the DSP blocks, $f$ is the clock frequency, and $\times2$ term accounts for multiply-accumulate (MAC) operations. The packing factor for DSP blocks, $p$, varies based on the bit-width of the operation, with $p = 1$ for 16-bit, $p = 2$ for 8-bit, and $p = 4$ for 4-bit multiply-accumulate operations.

Furthermore, the performance of an FPGA-based accelerator is also limited by the memory, which is related to the memory bandwidth (BW) and computation-to-communication (CTC) ratio:
\begin{equation}
    Peak\ memory\ bandwidth = BW \times CTC \ ratio.
\end{equation}

\begin{table}[!htbp]
    \centering
    \caption{Comparison between GPUs and FPGAs. Both V100 and U280 are compared using the PCIe version. Performance is the theoretical peak extracted from corresponding product datasheet~\cite{nvidia-v100,fpga-u280}.}
    \resizebox{1.0\linewidth}{!}{
    \begin{tabular}{c|cc}
    \toprule
        \textbf{Devices} & \textbf{V100 GPU}   & \textbf{Alveo U280 FPGA}  \\ \midrule
        Technology       & 12nm                & 16nm                      \\ \grayline
        Clock            & 1530MHz             & 300MHz                    \\ \grayline
        \multirow{2}{*}{Compute cores} & 5120 CUDA cores & \multirow{2}{*}{9024 DSP48E2} \\
        ~                & 640 Tensor cores    & ~                         \\ \grayline
        \multirow{2}{*}{Performance} & 14TFLOPs(FP32 CUDA) & \multirow{2}{*}{\textcolor{red}{24.5 TOPs(INT8)}} \\ 
        ~                & \textcolor{dkgreen}{112TFLOPs(FP16 Tensor)} & ~ \\ \grayline
        \multirow{2}{*}{Memory} & \multirow{2}{*}{32GB HBM2} & 32GB DDR4   \\ 
        ~                & ~                   & 8GB HBM2                  \\ \grayline
        \multirow{2}{*}{Bandwidth} & \multirow{2}{*}{900GB/s} & 38GB/s (DDR4)   \\ 
        ~                & ~                   & 460GB/s (HBM2)                  \\ \grayline
        \multirow{2}{*}{Power} & \multirow{2}{*}{250W} &  225W(Max)        \\ 
        ~                & ~                   & 100W(Typical)             \\ \grayline
        Price            & \$11,458            & \$7,717                   \\
        \bottomrule
    \end{tabular}
    } 
    
    \label{tab:comparision}
\end{table}

Table~\ref{tab:comparision} summarizes the major differences between GPUs and FPGAs, such as clock frequency, number of compute cores, and memory bandwidth. The significant difference in clock frequency contributes to a notable performance gap between GPUs and FPGAs. Even with optimization such as pruning and quantization~\cite{han2015deep_compression}, FPGA inference speed generally remains inferior to that of GPUs.

\begin{figure}[htbp]
    \centering
    \includegraphics[width=0.85\linewidth]{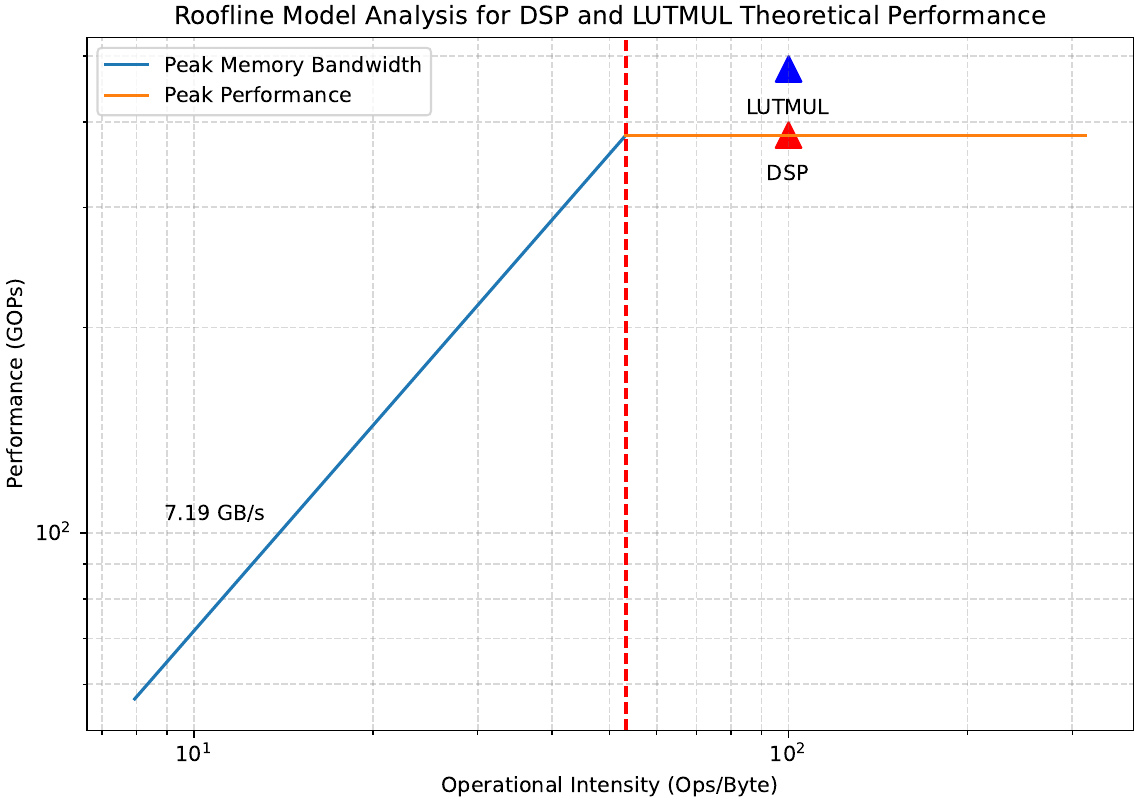}
    \caption{Roofline model analysis for \N{} and other DSP-based architectures. We take $\frac{1}{64}$ resource and memory bandwidth of U280 for analysis.}
    \label{fig:roofline}
\end{figure}

Figure~\ref{fig:roofline} shows the roofline model for U280. We take $\frac{1}{64}$ resource and HBM bandwidth of U280 for analysis. Conventional DSP-based accelerators are compute bound when the arithmetic intensity satisfies a threshold. Our \texttt{\N{}} exploits the potential of LUTs and can achieve higher parallelism by our LUT efficient mapping.

\subsection{Dataflow Architecture}

Dataflow architecture contrasts with the traditional control flow architecture. The dataflow nodes or processing elements can immediately start and execute when all its inputs are ready. Dataflow architecture employs simple operations, such as broadcast (one-to-many), map (element-wise, \eg{} activation function), zip (multi-operands, \eg{} convolution and matrix multiplication), and reduce (many-to-one, \eg{} pooling)~\cite{sambanova}. A key advantage of reconfigurable dataflow architecture is its ability to allow data to flow through the computation graph efficiently, significantly enhancing parallelism and minimizing memory access time.

\subsection{FPGA-based Neural Network Accelerator Architecture}

Since FPGAs have relatively limited on-chip resources, most of the FPGA accelerators~\cite{wu2019high, yu2020light, yan2021fpga, chang2021mix, sun2022film} map computation onto hardware and reuse the PE array.
Notably, ~\cite{lit2022auto} explores the intra-layer mixed-scheme quantization and maps vision transformer layers onto a General Matrix Multiply (GEMM) kernel, where each layer maintains a fixed ratio of this quantization scheme.
Systolic array architecture~\cite{wei2017automated} presents another method for efficiently mapping convolution and matrix multiplication onto FPGAs with high throughput.

FINN~\cite{umuroglu2017finn, blott2018finn} uses a dataflow architecture and integrates all layers of the network into a single FPGA. The resources for each layer can be adjusted according to computation requirements so that all layers are balanced and pipelined for better throughput. FINN is particularly well-suited for exploiting inter-layer quantization for neural networks because each layer has dedicated computation and memory resources.

\section{Algorithm-Hardware Co-Design for \textbf{\N{}}}
\label{sec:methods}

\subsection{Motivation}

The roofline model reveals a theoretical peak performance for DSP-based accelerators, applicable across various architectures such as GEMM, systolic array, or dataflow architecture. We can leverage LUT resources to perform multiplication and full parallelism to enable FPGA with greater performance. Given that the availability of LUTs usually outnumbers DSPs, using LUTs can potentially exceed the upper bound of performance of current DSP-based FPGA accelerators. 

\begin{figure}[htbp]
    \centering
    \includegraphics[width=0.9\linewidth]{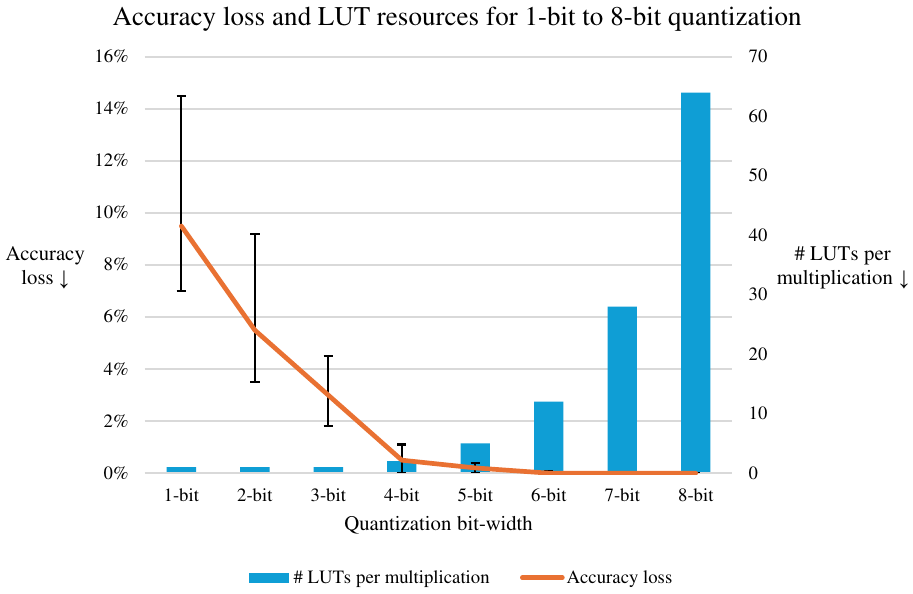}
    \caption{Accuracy loss and LUT resources for 1-bit to 8-bit quantization.}
    \label{fig:accuracy_resource}
\end{figure}

Figure~\ref{fig:accuracy_resource} shows the quantized neural network accuracy~\cite{jin2020adabits, yang2021fracbits} and the number of LUTs needed per multiplication by our method. We trade-off between accuracy and LUT usage and choose 4-bit as our quantization bit-width. Binary and ternary neural networks incur large accuracy drops and consume half of the LUTs that 4-bit uses as the output bits of LUTs are limited. Compared with higher bit-width quantization, 4-bit uses significantly fewer LUTs and has negligible accuracy loss.

\subsection{\textbf{\N{}} Design Flow}

\begin{figure}[htbp]
    \centering
    \includegraphics[width=1.0\linewidth]{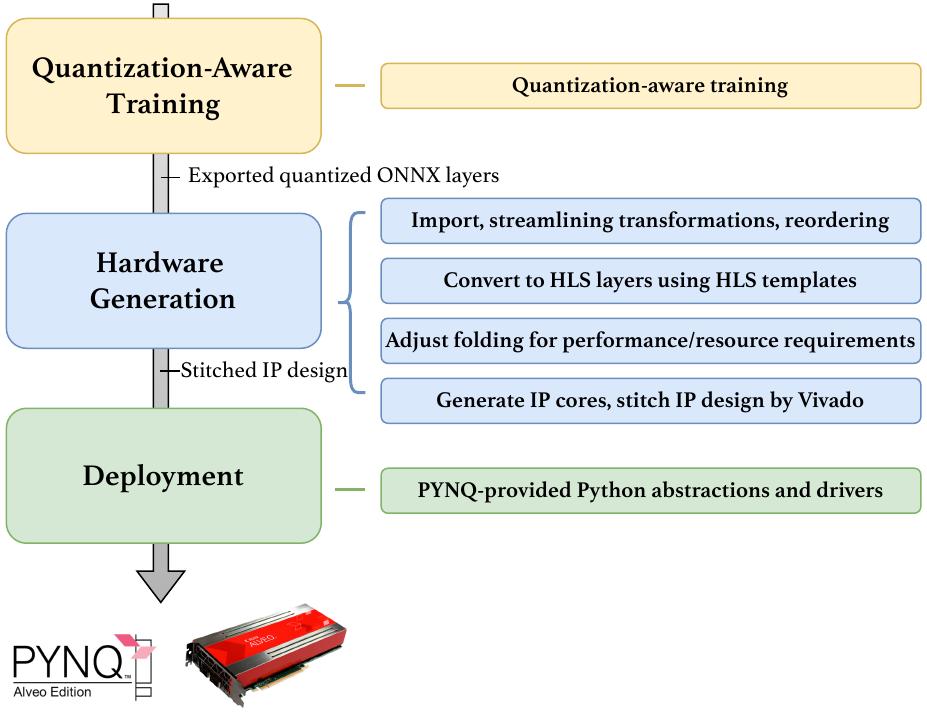}
    \caption{\textbf{\N{}} Design flow.}
    \label{fig:design_flow}
\end{figure}

Figure~\ref{fig:design_flow} depicts the \texttt{\N{}} design flow. Initially, we train the neural network in our quantization-aware training framework. The quantization bit-widths for weights and activations are adjustable hyperparameters. The final quantized neural network is exported in Open Neural Network Exchange (ONNX) format, facilitating subsequent hardware generation.

The ONNX intermediate representation is interpreted as a computation graph and undergoes a streamlining transformation~\cite{umuroglu2017streamlined}. The scaling factors of each channel and batch normalization layer are reordered and absorbed into the activation function, transforming into a multi-threshold unit. Each computation node is converted to a High-Level Synthesis (HLS) layer using our HLS templates. These HLS layers are folded according to performance and resource requirements and interconnected sequentially. The final hardware bitstream, generated by Vivado, is deployed on FPGA boards via the PYNQ framework.

\subsection{Reconfigurable Dataflow Architecture}

\begin{figure}[htbp]
    \centering
    \includegraphics[width=0.9\linewidth]{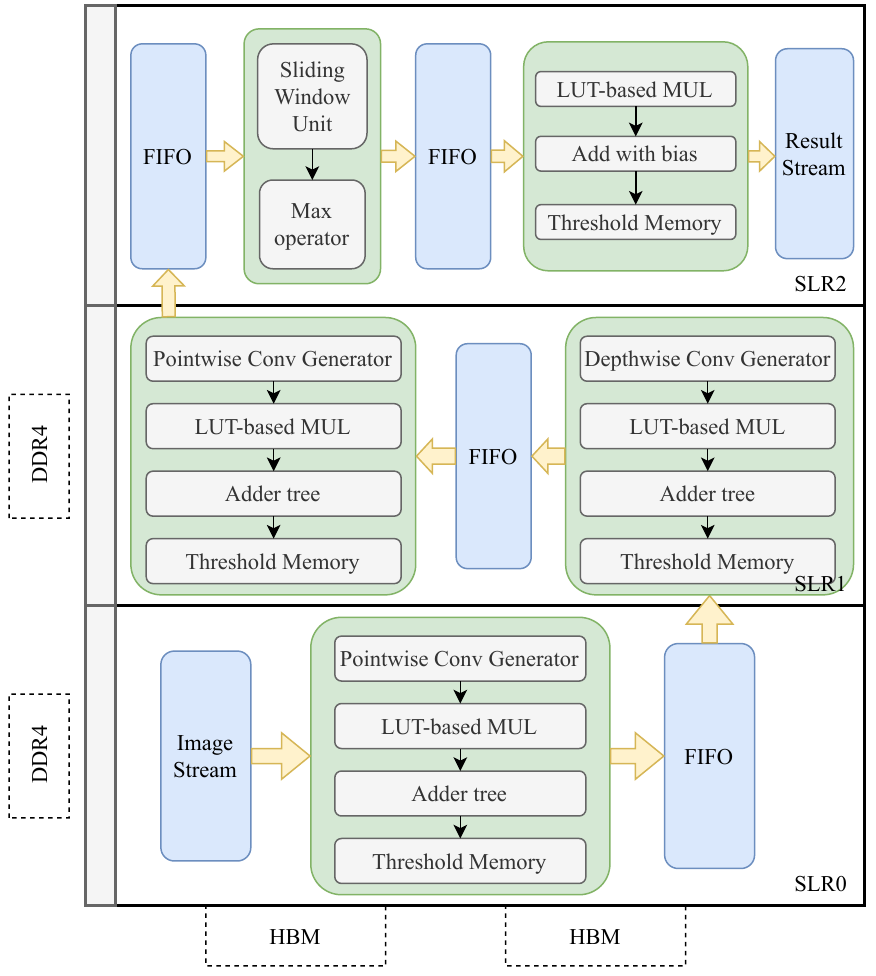}
    \caption{Hardware architecture of accelerator generated by \texttt{\N{}}. Our design is fully on-chip and does not use DRAM or HBM memory.}
    \label{fig:hardware}
\end{figure}

Figure~\ref{fig:hardware} illustrates the hardware architecture of a MobileNetV2~\cite{sandler2018mobilenetv2} implementation. This design, focusing on inverted residual blocks, employs a First In, First Out (FIFO) buffer between layers to store activations. The architecture uses a reconfigurable dataflow architecture.

Our design spans all Super Logic Regions (SLRs) to maximize hardware resources. Signals only traverse SLRs when the current SLR resources are insufficient for the next layer to avoid severe timing violations. Dataflow architecture is inherently suited for design spanning multiple SLRs and can be scaled up, enabling additional FPGAs connected via network for deploying larger networks~\cite{diaconu2023machine}.

\subsection{Convolution Generator}
\label{subsec:conv-generator}

For convolutional layers, the convolution operations can be lowered to matrix-matrix multiplications. These can be mapped in a streaming manner and fed to the multiplication kernel. The multiplication kernel is fully paralleled to perform a matrix-vector multiplication, where the weights are stationary vectors and activations are streaming inputs. Therefore, we need a convolution generator to perform the im2col operations: reading data from FIFO, moving across input images to form an image matrix, and streaming the output to the multiplication kernel.

The convolution generator accommodates various configurations, including pointwise, depthwise, and standard convolution with different kernel sizes and strides, since each kind of convolutional layer expects different input data sequences, necessitating specific generator settings.

\subsection{Look-Up Table based Efficient Multiplication}
\label{subsec:lutmul}

\begin{figure*}[htbp]
    \centering
    \begin{subfigure}[h]{0.55\linewidth}
        \centering
        \includegraphics[width=0.9\linewidth]{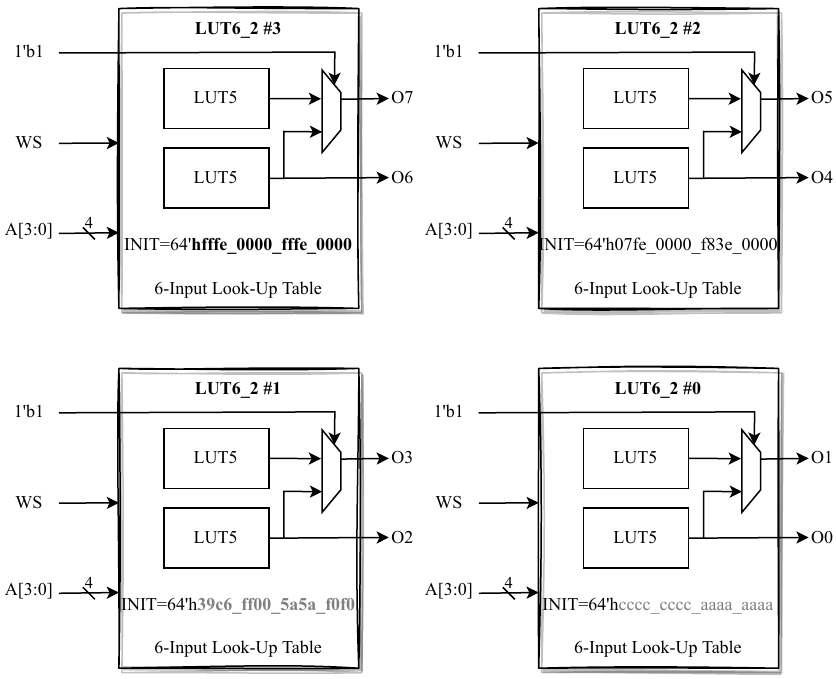}
    \end{subfigure}
    \hfill 
    \begin{subfigure}[h]{0.42\linewidth}
    \centering
    \resizebox{0.88\linewidth}{!}{
    \begin{tabular}{ccc}
    \toprule
    \multirow{2}{*}{\textbf{Activation}} & \multicolumn{2}{c}{\textbf{Multiplication results}} \\ 
    \cmidrule{2-3}
    & \text{Weight=1, WS=0} & \text{Weight=-3, WS=1} \\ \midrule
    4'b0000       & \textbf{0 0 }\text{0 0 }\textbf{\textcolor{Azure4}{0 0 }}\textcolor{Azure4}{0 0} 
                  & \textbf{0 0 }\text{0 0 }\textbf{\textcolor{Azure4}{0 0 }}\textcolor{Azure4}{0 0} \\
    4'b0001       & \textbf{0 0 }\text{0 0 }\textbf{\textcolor{Azure4}{0 0 }}\textcolor{Azure4}{0 1} 
                  & \textbf{1 1 }\text{1 1 }\textbf{\textcolor{Azure4}{1 1 }}\textcolor{Azure4}{0 1} \\
    4'b0010       & \textbf{0 0 }\text{0 0 }\textbf{\textcolor{Azure4}{0 0 }}\textcolor{Azure4}{1 0} 
                  & \textbf{1 1 }\text{1 1 }\textbf{\textcolor{Azure4}{1 0 }}\textcolor{Azure4}{1 0} \\
    4'b0011       & \textbf{0 0 }\text{0 0 }\textbf{\textcolor{Azure4}{0 0 }}\textcolor{Azure4}{1 1} 
                  & \textbf{1 1 }\text{1 1 }\textbf{\textcolor{Azure4}{0 1 }}\textcolor{Azure4}{1 1} \\
    4'b0100       & \textbf{0 0 }\text{0 0 }\textbf{\textcolor{Azure4}{0 1 }}\textcolor{Azure4}{0 0} 
                  & \textbf{1 1 }\text{1 1 }\textbf{\textcolor{Azure4}{0 1 }}\textcolor{Azure4}{0 0} \\
    4'b0101       & \textbf{0 0 }\text{0 0 }\textbf{\textcolor{Azure4}{0 1 }}\textcolor{Azure4}{0 1} 
                  & \textbf{1 1 }\text{1 1 }\textbf{\textcolor{Azure4}{0 0 }}\textcolor{Azure4}{0 1} \\
    4'b0110       & \textbf{0 0 }\text{0 0 }\textbf{\textcolor{Azure4}{0 1 }}\textcolor{Azure4}{1 0} 
                  & \textbf{1 1 }\text{1 0 }\textbf{\textcolor{Azure4}{1 1 }}\textcolor{Azure4}{1 0} \\
    4'b0111       & \textbf{0 0 }\text{0 0 }\textbf{\textcolor{Azure4}{0 1 }}\textcolor{Azure4}{1 1} 
                  & \textbf{1 1 }\text{1 0 }\textbf{\textcolor{Azure4}{1 0 }}\textcolor{Azure4}{1 1} \\
    4'b1000       & \textbf{0 0 }\text{0 0 }\textbf{\textcolor{Azure4}{1 0 }}\textcolor{Azure4}{0 0} 
                  & \textbf{1 1 }\text{1 0 }\textbf{\textcolor{Azure4}{1 0 }}\textcolor{Azure4}{0 0} \\
    4'b1001       & \textbf{0 0 }\text{0 0 }\textbf{\textcolor{Azure4}{1 0 }}\textcolor{Azure4}{0 1} 
                  & \textbf{1 1 }\text{1 0 }\textbf{\textcolor{Azure4}{0 1 }}\textcolor{Azure4}{0 1} \\
    4'b1010       & \textbf{0 0 }\text{0 0 }\textbf{\textcolor{Azure4}{1 0 }}\textcolor{Azure4}{1 0} 
                  & \textbf{1 1 }\text{1 0 }\textbf{\textcolor{Azure4}{0 0 }}\textcolor{Azure4}{1 0} \\
    4'b1011       & \textbf{0 0 }\text{0 0 }\textbf{\textcolor{Azure4}{1 0 }}\textcolor{Azure4}{1 1} 
                  & \textbf{1 1 }\text{0 1 }\textbf{\textcolor{Azure4}{1 1 }}\textcolor{Azure4}{1 1} \\
    4'b1100       & \textbf{0 0 }\text{0 0 }\textbf{\textcolor{Azure4}{1 1 }}\textcolor{Azure4}{0 0} 
                  & \textbf{1 1 }\text{0 1 }\textbf{\textcolor{Azure4}{1 1 }}\textcolor{Azure4}{0 0} \\
    4'b1101       & \textbf{0 0 }\text{0 0 }\textbf{\textcolor{Azure4}{1 1 }}\textcolor{Azure4}{0 1} 
                  & \textbf{1 1 }\text{0 1 }\textbf{\textcolor{Azure4}{1 0 }}\textcolor{Azure4}{0 1} \\
    4'b1110       & \textbf{0 0 }\text{0 0 }\textbf{\textcolor{Azure4}{1 1 }}\textcolor{Azure4}{1 0} 
                  & \textbf{1 1 }\text{0 1 }\textbf{\textcolor{Azure4}{0 1 }}\textcolor{Azure4}{1 0} \\
    4'b1111       & \textbf{0 0 }\text{0 0 }\textbf{\textcolor{Azure4}{1 1 }}\textcolor{Azure4}{1 1} 
                  & \textbf{1 1 }\text{0 1 }\textbf{\textcolor{Azure4}{0 0 }}\textcolor{Azure4}{1 1} \\
    \bottomrule
    \end{tabular}
    } 
    \end{subfigure}
    \caption{Illustration of \textbf{\N{}} for efficient multiplication via look-up tables. The left-hand side figure demonstrates how to use LUT6\_2 primitive for embedding multiplication results of weights and input activations. The right-hand side table demonstrates the multiplication results of two example weights and how to generate the corresponding look-up table contents. \normalfont{The weights (int4) and multiplication output (int8) are using two's complement representation, while activation are all unsigned numbers (uint4). The Most Significant Bit (MSB) of LUT6\_2 input is configured as `1' to enable two output ports. The bit below the MSB is a Weight Select (WS) signal to select between two weights. The lowest 4-bit inputs serve as activation inputs. Our method embeds two int4 weights inside four LUT6, a resource-efficient approach contrasting with the LUT6-instantiated general multipliers, which consume 6-14$\times$ more LUT6 resources. Two used example weights are 1 and -3 respectively. The embedded LUT contents for these four LUTs are \textbf{64'hfffe\_0000\_fffe\_0000}, \text{64'h07fe\_0000\_f83e\_0000}, \textbf{\textcolor{Azure4}{64'h39c6\_ff00\_5a5a\_f0f0}}, and \textcolor{Azure4}{64'hcccc\_cccc\_aaaa\_aaaa}, respectively.}}
    \label{fig:lut_contents}
\end{figure*}

Figure~\ref{fig:lut_contents} demonstrates the look-up table based multiplication kernels and how to determine look-up table initialization (INIT) values. After embedding the weights into look-up tables, our look-up tables transform into efficient constant multipliers~\cite{hardieck2019reconfigurable}. Our look-up table based multiplier is efficient in resources, utilizing only 2 LUTs on average for a single 4-bit multiplication, compared with a general multiplier that consumes 13-28 LUTs for an equivalent operation. The choice of 4-bit quantization is pivotal as it maintains model accuracy and optimizes look-up table usage, as shown in Figure~\ref{fig:accuracy_resource}. We show the number of LUT6 (6:1 LUT, 6-bit input, 1-bit output) for a general n-bit multiplication (n:2n LUT, n-bit input, 2n-bit output) using our method:
\begin{equation}
    \#LUTs = \frac{2n \times 2^{n}}{1 \times 2^6}.
\end{equation}

Algorithm~\ref{alg:lutmul-kernel} shows the pseudo High-Level Synthesis (HLS) code for look-up table based multiplication. The table contents are derived from pre-computed weights. The weights of convolutional layers are fully paralleled, meaning that the $COUT$ channel in Algorithm~\ref{alg:lutmul-kernel} refers to the output channels, and the $CIN$ channel refers to the input channels times the kernel size squared. These dimensions (four in total) are fully unrolled in the spatial domain. The remaining input feature map height and width dimensions are pipelined in the temporal domain. Input activations are streamed from the convolution generator and passed through look-up tables. The output results are multiplication results. They are accumulated, go through the threshold unit, and generate activations for the next layer.

\begin{algorithm}[htbp]
\small
\caption{Look-up table based multiplication kernel}
\label{alg:lutmul-kernel}
\begin{algorithmic}[1]
\REQUIRE Streaming parallel input from the Convolution Generator and pre-computed look-up table contents
\ENSURE Streaming output for the next layer
\FOR {$i \gets 1 \ to \ ROWS \times COLS $}
\STATE \hspace{-1\algorithmicindent}\#pragma HLS PIPELINE II=1
\STATE {$input \gets src.read()$}
\FOR {$co \gets 1 \ to \ COUT$}
\STATE \hspace{-2\algorithmicindent}\#pragma HLS UNROLL
\FOR {$ci \gets 1 \ to \ CIN$}
\STATE \hspace{-3\algorithmicindent}\#pragma HLS UNROLL
\STATE $mul[co][ci] \gets lut[co][ci][input[ci]]$
\ENDFOR
\ENDFOR
\FOR {$co \gets 1 \ to \ COUT$}
\STATE \hspace{-2\algorithmicindent}\#pragma HLS UNROLL
\FOR {$ci \gets 1 \ to \ CIN$}
\STATE \hspace{-3\algorithmicindent}\#pragma HLS UNROLL
\STATE $output[co] += mul[co][ci]$
\ENDFOR
\ENDFOR
\STATE {$dst.write(output)$}
\ENDFOR
\end{algorithmic}
\end{algorithm}

\subsection{Quantization-Aware Training}

Quantization~\cite{han2015deep_compression} and DSP packing~\cite{sun2022film} have become a standard approach for mapping neural networks onto FPGA-based accelerators, as FPGAs' LUTs and DSP blocks are not optimized for floating-point but ideal for integer or fixed-point operations. Quantization, paired with DSP packing, reduces resource demands for the multiplications and improves throughput.

The quantization operation is defined as:
\begin{equation}
    y = quantize(x) = clamp(round(\frac{x}{s}+z), y_{min}, y_{max}),
\end{equation}
where $x$ is the floating-point value to quantize, $s$ is the scaling factor of the output quantized tensor, and $z$ is the zero-point or quantization bias coefficient. The function, $round$, can be round-to-even or round-to-zero, and $clamp$ performs clipping inclusive of the boundaries $y_{min}$ and $y_{max}$.

For the reverse process, to compute the floating-point representation of a quantized value, we define the dequantize operation as:
\begin{equation}
    dequantize(y) = s(y-z),
\end{equation}
where $y$ is a quantized tensor, $z$ is its zero-point, and $s$ is the scaling factor.

Quantization introduces errors to the trained model parameters and results in performance degradation. Quantization-Aware Training (QAT) is a popular approach that retrains the model with quantized parameters on the pretraining dataset to converge to the pretrained model performance. The usual forward and backward passes are performed on the quantized model in floating point, and the model parameters are quantized after each gradient update. In particular, it is important to do this projection after the weight update is performed in floating point precision. Performing the backward pass with floating point is vital, as accumulating the gradients in quantized precision can result in zero gradients or gradients with high error, especially in low-precision scenarios~\cite{gholami2022survey}.

\section{Evaluation}
\label{sec:experiments}

\begin{table*}[htbp]
\tabcolsep 2pt
\caption{Comparisons of MobileNet implementations between previous FPGA-based accelerators.}
\centering
\resizebox{0.9\linewidth}{!}{
\begin{tabular}{c|c|c|c|c|c|c|c}
\toprule
\multirow{2}{*}{Implementation} & FINN & FPL'19 & Light-OPU & FPL'21 & Mix \& Match & FILM-QNN & \bf{\N{}}\\
~ & ~\cite{blott2018finn} & ~\cite{wu2019high} & ~\cite{yu2020light} & ~\cite{yan2021fpga} & ~\cite{chang2021mix} & ~\cite{sun2022film}  & (Ours) \\
\midrule
 Network & MobileNetV1& MobileNetV2& MobileNetV3& MobileNetV2& MobileNetV2& MobileNetV2& MobileNetV2\\ \hline
 Bit-Width & W4A4 & W8A8 & W8A8 & W8A8 & W4A4 & W8A5\&W4A5 & W4A4\\ \hline
 Top-1 Accuracy & 70.4\% & 68.1\% & 66.7\% & 70.8\% & 65.6\% & 65.7\% & \textbf{70.95\%}\\ \hline
 Platform & Alveo U280 & ZU9EG & XC7K325T & XC7V690T& XC7Z045& ZU9EG&Alveo U280\\ \hline
 Frequency (MHz) & 333& 333& 200& 150& 100& 150&333\\ \hline
 LUT& 501363& 161944& 173522& 308449& 145049&  180100&529242\\
 FF& 476316& 301416& 241175& 278926& 111575& -&503192\\
 BRAM36& 898& 771& 193.5& 941.5& 225.5& 440.5&1119\\
 DSP& 106& 2070& 704& 2160& 900& 2092&106\\
\hline
Power (W) & 41.69& -& 8.5*& 11.35& -& 12.9&42.12\\
\hline
Frame Rate (FPS) & 925& 809.8& 332.6& 302.3& 549.3& 537.9&\textbf{1627}\\
Throughput (GOPS) & 556.4& 487.1& 84.48*& 181.8& 326.9& 320.1&\textbf{978.6}\\
\hline
Energy Efficiency (GOPS/W) & 13.35& -& 9.9& 16.02& -& \textbf{24.8}&23.23\\
\bottomrule
\end{tabular}
} 

{\vspace{0.15cm} \footnotesize \raggedright \hspace{1cm} Note: `-' means that the result is not given in the original publications, and `*' means that the result is inferred from the original publications. \par}
\label{tab:experiments}
\end{table*}

\subsection{Experimental Setup}
To evaluate the performance of \texttt{\N{}}, we implement MobileNetV2 on FPGAs and compare it with existing FPGA-based MobileNet accelerators. MobilenetV2~\cite{sandler2018mobilenetv2} has 3.4M parameters and achieves 71.88\% top-1 accuracy on the ImageNet dataset~\cite{deng2009imagenet}. We utilize the FINN framework~\cite{blott2018finn} as our foundational platform.
For quantization, we adopt PyTorch 1.13.0 and Brevitas 0.9.1~\cite{brevitas}. Specifically, we choose 4-bit for weights and activations quantization except for the first and last layers, which are set as 8-bit. To preserve the accuracy of the MobileNetV2 model, we apply the channel-wise quantization scheme for our model. Our quantized MobileNetV2 network is trained for 420 epochs, culminating in a 70.95\% top-1 accuracy evaluated on the ImageNet dataset~\cite{deng2009imagenet}.

For the hardware evaluation, the utilized development platform is the AMD Xilinx Alveo U280 data center accelerator card on the Open Cloud Testbed (OCT)~\cite{zink2021open}. We implement the first 15 layers of MobileNetV2 in a fully parallel manner and fold the remaining layers for resource optimization. To maximize the computation efficiency without timing violation, the working frequency is set to 333 MHz for all the designs implemented through Vitis HLS/Vivado 2022.1.

\subsection{Experimental Results}

Table~\ref{tab:experiments} showcases the hardware performance and comparisons with other FPGA-based MobileNet accelerators. Most of these accelerators are tailored for edge FPGAs, such as ZU9EG, except for FINN, which has data center accelerator implementation for MobileNetV1. The FINN result is generated and tested on the same device as our implementation, while other data points are extracted from their original publications.

In terms of accuracy, our model achieves the best 70.95\% top-1 accuracy on ImageNet among all implementations. Quantization-aware training effectively mitigates quantization errors, preserving the model original accuracy, even with 4-bit quantized weights and activations.

As for the inference performance, our implementation achieves a throughput of 1627 images per second. Our implementation consumes the most FPGA resources but could still fit on a single Alveo U280. However, it is noteworthy that our implementation also yields a 23.23 GOPS/W energy efficiency, marginally lower than the FLIM-QNN~\cite{sun2022film}, which is implemented on a more power-efficient edge FPGA board.

\subsection{Discussion}

\begin{figure}[htbp]
    \centering
    \includegraphics[width=0.9\linewidth]{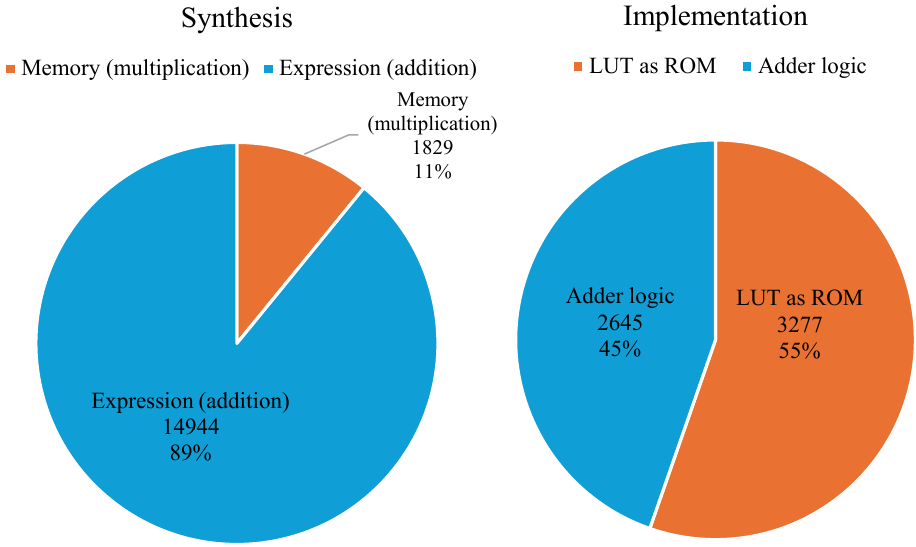}
    \caption{LUT Resource breakdown of the second convolution layer in MobileNetV2 using \textbf{\N{}}.}
    \label{fig:resource_breakdown}
\end{figure}
Figure~\ref{fig:resource_breakdown} shows the LUT resource breakdown of the second convolution layer in MobileNetV2 using \texttt{\N{}}, which is a $1 \times 1$ convolution kernel and has 32 input channels and 32 output channels. For these 1024 4-bit weights, multiplication operations use 1829 LUTs after HLS synthesis, which matches the theoretical analysis of \texttt{\N{}}. However, HLS instantiates an adder for each addition operation to achieve an II of 1, resulting in a high usage of LUT for adder logic. After Vivado implementation, the LUT usage decreased to 5922. Vivado optimizes the logic and instantiates 3277 LUTs as ROM and 2645 LUTs as adder and other logic. Even though adder logic accounts for a large part of resources, the parallel MAC performance by \texttt{\N{}} still outperforms the DSP packing method using the same number of resources. 

\subsection{Comparisons with Related Works}

Our method is not only limited to integer multiplication, but can also be extended to customized data formats, such as FP4 and MXFP4~\cite{rouhani2023microscaling}, while DSP packing is designed efficiently for integer formats. LUTNet~\cite{wang2019lutnet, wang2022logic} also utilizes LUT for inference and explores the flexibility of LUT. However, LUTNet design suffers from low accuracy when the network becomes larger. PolyLUT~\cite{andronic2023polylut} trains multivariate polynomials instead of linear functions and embeds piecewise polynomial functions into LUTs. CompressedLUT~\cite{khataei2024compressedlut} proposes a lossless LUT compression method and is efficient for non-linear functions and large LUT logic blocks, such as~\cite{umuroglu2020logicnets, hong2023algorithms, gerlinghoff2024table}. Our method maps MAC operations to the single-LUT level, and Vivado can handle remaining logic optimization efficiently. 
\section{Conclusion}

We propose \texttt{\N{}}, an efficient method that leverages look-up tables for multiplication in convolutional neural networks. Compared with the general multiplier, our method is efficient in resources, which only needs two look-up tables on average for a single 4-bit multiplication. Compared with other DSP-based FPGA accelerators, \texttt{\N{}}'s reconfigurable dataflow architecture enables full parallelism, reduces memory access time, and increases the theoretical upper bound of performance. Experimental results demonstrate that our design maintains a top-1 accuracy of 70.95\% on the ImageNet dataset and achieves a throughput of 1627 images per second on a single Alveo U280 FPGA, outperforming other FPGA-based MobileNet accelerators. 

\begin{acks}
This research was supported in part by the National Science Foundation under Grants CCF-1901378, CNS-1925658, and CNS-2319962.
\end{acks}

\bibliographystyle{ACM-Reference-Format}
\bibliography{references}


\end{document}